\documentclass[12pt]{article}
\usepackage{amssymb,amsmath}

\def\beq{\begin{equation}}
\def\eeq{\end{equation}}
\def\br{\begin{eqnarray}}
\def\er{\end{eqnarray}}
\def\benu{\begin{enumerate}}
\def\eenu{\end{enumerate}}
\def\nn{\nonumber}

\def\MPl{M_{_{\rm Pl}}^2}
\def\lpl{l_{_{\rm Pl}}^2}
\def\MPlh{M_{_{\rm Pl}}}
\def\MPlD{M_{_{\rm P}}^{(D - 2)}}

\def\SBH{{\mathcal S}_{\rm BH}}
\newcommand{\SEN}{S_{_{\rm ent}}}
\def\AH{{\cal A}_{\rm H}}
\def\AHD{{\cal A}_{\rm D}}
\begin{document}

\begin{titlepage}
\pagestyle{empty}
\baselineskip=21pt
\vspace{2cm}
\begin{center}
{\bf {\Large 
Corrections to Bekenstein-Hawking entropy --- Quantum or not-so quantum?
}}
\end{center}
\begin{center}
\vskip 0.2in
{\bf S. Shankaranarayanan}
\vskip 0.1in
{\it School of Physics,  
Indian Institute of Science Education and Research, \\
CET campus, Trivandrum 695016, India} \\
{\tt Email: shanki@iisertvm.ac.in}\\
\end{center}

\vspace*{0.5cm}

\begin{abstract}
Hawking radiation and Bekenstein--Hawking entropy are the
two robust predictions of a yet unknown quantum theory of gravity.
Any theory which fails to reproduce these predictions is certainly
incorrect. While several approaches lead to Bekenstein--Hawking
entropy, they all lead to different sub-leading corrections. In this
article, we ask a question that is relevant for any approach: Using
simple techniques, can we know whether an approach contains quantum
or semi-classical degrees of freedom? Using naive dimensional
analysis, we show that the semi-classical black-hole entropy has the
same dimensional dependence as the gravity action. Among others,
this provides a plausible explanation for the connection between
Einstein's equations and thermodynamic equation of state, and that
the quantum corrections should have a different scaling behavior.
\end{abstract}

\end{titlepage}

\baselineskip=18pt

Entropy is a derived quantity and does not show up in any
fundamental equation of motion. However, in any physical theory,
entropy takes unique position amongst other quantities. This is due
to the fact that entropy relates the macroscopic and microscopic
degrees of freedom (DOF) through Boltzmann
relation~\cite{1978-Wehrl-RMP}:
\begin{equation}
S = k_B \, \ln \Omega
\end{equation}
where $k_B$ is Boltzmann constant and $\Omega$ is total number of
micro-states. Hence, it is not surprising that there has been
intense research activity in obtaining the microscopic description
of Bekenstein--Hawking entropy \cite{1973-Bekenstein-PRD}:
\begin{equation}
\SBH = \frac{k_{_B}}{4} \frac{\AH}{\lpl} =
\frac{k_{_B}}{4} \, \MPl \, \AH
\label{eq:SBH}
\end{equation}
where, $\AH$ is the area of black-hole horizon, $\MPl \equiv 1/(8
\pi G)$, $\lpl$ are Planck mass and Planck length, respectively.

While several approaches lead to $\SBH$, none of these approaches
can be considered to be complete. For instance, in string
computations, BPS states are well-defined only for (near) extremal
black-holes \cite{1996-Strominger.Vafa-PLB,2008-Sen-GRG}.  In
conformal field theory approach
\cite{1999-Carlip-CQG,2000-Carlip-CQG}, where the horizon is treated
as boundary, the vector fields (which generate the symmetries) do
not have a well-defined limit at the horizon
\cite{2001-Dreyer.etal-CQG}. Besides, all these approaches lead to
different sub-leading corrections to $\SBH$. For instance, conformal
field theory \cite{2000-Carlip-CQG} and quantum geometry approaches
\cite{2000-Kaul.Majumdar-PRL} lead to logarithmic corrections while
the string \cite{2008-Sen-GRG} and entanglement computations
\cite{2008-Das.etal-PRD,2008-Das.etal-Arx} lead to power-law
corrections.

In hindsight, one can say this is probably expected; different
approaches count different microscopic states that are valid in
domains of their applicability.  In the absence of a consistent
quantum theory of gravity, it is not possible to know the
microscopic DOF, and hence, the subject of black-hole thermodynamics
resembles a {\it jig-saw puzzle}. To put several pieces together in
this puzzle, the best strategy is to slowly build a coherent picture
and hope to understand/solve some of these problems. The purpose of
this article is an attempt in this direction and we ask: Is there a
way one can classify these different subleading corrections to the
Bekenstein--Hawking entropy? In other words, {\it Using simple
techniques, can we know whether an approach contains semi-classical
or quantum DOF?} The answer to this question is relevant for any
approach to black-hole entropy.

We show that the naive dimensional analysis \cite{dim-analysis}
provides crucial information about the semi-classical and(or)
quantum nature of the sub-leading terms. Among others, this provides
an explanation for the connection between Einstein's equations and
thermodynamic equation of state. Using this, we argue that quantum
entanglement is crucial to gain insights on black-hole
thermodynamics.

But, why is it important to understand sub-leading corrections to
$\SBH$?  In order to exemplify this, let us compare black-hole
entropy with ideal gas entropy. The classical entropy of mono-atomic
ideal gas is given by
\begin{equation}
\frac{S_{_{\rm ideal}}}{k_{_B} \, N}  = \ln \left( V \, T^{3/2} \right) 
\label{eq:ideal-class}
\end{equation}
where $V, T, N$ correspond to volume, temperature and number of
particles, respectively.

Assuming that all atoms move independently, we can obtain the number
of quantum states and, hence, the Sackur--Tetrode entropy
\cite{1996-Pathria-Statisticalmechanics}
\begin{equation}
\frac{S_{_{\rm ST}}}{k_{_B} \, N} =
\ln \left(V \, T^{3/2} \right)
+ \frac{1}{2} \ln \frac{M^{3}}{N^{5}}
+  \frac{3}{2} \ln \left( \frac{4 \pi k_{_B}}{3 \hbar^2 } \right) \, ,
\label{eq:ideal-ST}
\end{equation}
where $M$ is the mass of the gas.

Among others, there are two main reasons for the relevance of
$S_{_{\rm ST}}$ to black-hole entropy:
Firstly, $S_{_{\rm ST}}$ depends on the mass of the DOF of an ideal
gas---{\it the individual atom}; the classical expression has no
explicit mass dependence.  In other words, varied DOF can lead to
identical classical expression while sub-leading terms which contain
information about the DOF will be different. Similarly, several
approaches to black-hole entropy lead to identical $\SBH$; however,
they lead to different sub-leading~corrections.

Secondly, quantum correction to classical entropy [third term in the
RHS of (\ref{eq:ideal-ST})] does not depend on the macroscopic
quantities.  It is needless to say that this could not have been
foreseen by physical arguments. In the same manner, it would be
impossible to predict quantum corrections to $\SBH$. If one uses
symmetry arguments based on classical action, then what we may
obtain, as discussed below, will be proportional to the form
obtained from dimensional analysis\cite{2010-Caravelli.Modesto-Arx}.

Having addressed the importance of sub-leading corrections, we show
that naive dimensional analysis provides information about the
quantum/semi-classical nature of the sub-leading terms. Let us
consider the Einstein--Hilbert action in 4-dimensions:
\begin{equation}
\label{eq:SEH}
S_{_{\rm EH}} = \frac{\MPl}{2} \int d^4 x \, \sqrt{-g} \, R  \, .
\end{equation}
Dimensional analysis of the above action leads to
\begin{equation}
S_{_{\rm EH}} \propto \MPlh^{2} \, \times \, [L]^{2}
\label{eq:EH-Ndim}
\end{equation}
Dimensional analysis of 4-dimensional Bekenstein--Hawking entropy
(\ref{eq:SBH}) leads to
\begin{equation}
\SBH \propto \MPlh^{2} \, \times \, [L]^{2}
\label{eq:BH-Ndim}
\end{equation}
Attentive reader might realize that the Einstein--Hilbert action
(\ref{eq:EH-Ndim}) and $\SBH$ (\ref{eq:BH-Ndim}) have the same
dimensional dependence, indicating that the semi-classical
black-hole entropy seem to follow the (classical) gravity action and
ask, does this relation hold for a general gravity action?

Let us now consider a D-dimensional gravity action:
\br
S_{_{\rm Gen}} &=& \frac{1}{16 \pi G_{_D}} \int d^D x \sqrt{-g}
\left[R + \alpha \, F(R^2) + \beta \, G(R^3) + \cdots \right]  \nn \\
\label{eq:SGen}
&=& \frac{\MPlD}{16 \pi} \int d^D x \sqrt{-g} \left[R + \alpha \,
 F(R^2) + \beta \, G(R^3) + \cdots \right]
\er
where $G_D, M_{_{\rm P}}$ are D-dimensional Newton's constant and
Planck mass, respectively, $F(R^2)$ include combination of $R^2$,
$R_{AB} R^{AB}$, $R_{ABCD} R^{ABCD}$ terms, $G(R^3)$ includes cubic
terms, and $\alpha, \beta$ are dimension-full constants. The above
action includes Lovelock gravity whose equations of motion are
quasi-linear
\cite{1971-Lovelock-JMP,2009-Kothawala.Padmanabhan-PRD}. Dimensional
analysis of this action leads to:
\begin{equation}
S_{_{\rm Gen}} \propto \MPlD \, \times \, [L]^{D - 2}
\left[1 + \frac{\alpha}{[L]^2}  + \frac{\beta}{[L]^4} + \cdots \right]
\label{eq:Gen-Ndim}
\end{equation}
The Noether charge entropy corresponding to this action is given by
\cite{1993-Wald-PRD}:
\begin{equation}
S_{_{\rm NC}} =  \MPlD \frac{\AHD}{4} \left[1 + \alpha \AHD^{2/(D - 2)}  +
\beta \AHD^{2/(D - 2)} + \cdots \right] \, ,
\label{eq:SNC}
\end{equation}
where $\AHD$ is the horizon area of black-holes in D-dimensional
space-time. It is important to note that in deriving
Bekenstein--Hawking and Noether charge entropy, it is assumed that
the back-reaction of the Hawking particles are negligible.

Dimensional analysis of the entropy leads to:
\begin{equation}
S_{_{\rm NC}} \propto \MPlD \, \times \, [L]^{D - 2}
\left[1 + \frac{\alpha}{[L]^2}  + \frac{\beta}{[L]^4} + \cdots \right]
\label{eq:NC-Ndim}
\end{equation}
This observation indicates that the (semi-classical)
black-hole---like $\SBH$ and No-ether charge---entropy in any
gravity theory follow the form of the classical gravity action. So,
what are the physical consequences of this observation?
Firstly, this feature is specific to gravity and, to author's
knowledge, can not be seen in other fundamental interactions. Comparing
Equation (\ref{eq:ideal-ST}) with the electromagnetic action:
\beq
S_{EM} = - \frac{1}{4} \int d^4 \, x \, F_{\mu\nu} F^{\mu\nu} \, ,
\eeq
it is clear that the entropy of ideal gas and the electromagnetic
action do not have same dimensional~dependence.

The above observation can be viewed as the primary reason as to why
Einstein's equations can be viewed as thermodynamic equation of
state \cite{1995-Jacobson-PRL,2008-Padmanabhan-Arx}.  The crucial
input, which leads to thermodynamic equation from Einstein
equations, is the form of the entropy (\ref{eq:SBH}, \ref{eq:SNC}).
For instance, if we consider the power-law corrections to the
Bekenstein--Hawking entropy in 4-dimensional space-times arising
from entanglement \cite{2008-Das.etal-PRD}, and use the approach of
Jacobson or Padmanabhan, Einstein equations {\it can not} be
rewritten as first law of thermodynamics. In the recent proposal by
Verlinde \cite{2010-Verlinde-Arx,2007-Makela-Arx}, interpreting
gravity as entropic force, there is an implicit assumption about the
entropy-area relation \cite{2010-Modesto.Randono-Arx}.

Secondly, in ideal gas, the quantum corrections to the
(semi)classical entropy do not have any volume dependence. For
black-hole entropy, this suggests that quantum gravitational
corrections to $\SBH$ will include terms which may not follow the
form of the classical gravity action.\ At least two of the
approaches to black-hole entropy do seem to agree with this
observation:  (i) In quantum geometry approach, it was shown that
Hilbert space of the horizon of spherically symmetric space-time is
$2d \, SU(2)_k$ Wess--Zumino model leading to generic logarithmic
corrections \cite{2000-Kaul.Majumdar-PRL}. (ii) Entanglement entropy
of the metric perturbations, about the black-hole background in
4-dimensional general relativity, lead to power-law corrections
\cite{2008-Das.etal-PRD,2008-Das.etal-Arx}.

Lastly, this provides a simple way to classify approaches which
predict corrections to Bekenstein--Hawking entropy. For instance,
approaches discussed in \cite{2008-Sen-GRG,1995-Demers.etal-PRD}
lead to the form which is similar to Equation (\ref{eq:SNC}) while
that obtained in \cite{2008-Das.etal-PRD} do not follow Noether
charge entropy. In other words, this suggests that if any approach
to black-hole entropy predicts the same dimensional form as the
classical action of gravity, then this approach {\it only} provides
semi-classical, and not quantum, structure of gravity. This might
seem a strong assertion, however, it would be an even stronger claim
if one says that the quantum corrections to $\SBH$ follow the same
dimensionality of the classical action. For instance, the power-law
corrections to the Bekenstein-Hawking entropy obtained by Demers et
al. \cite{1995-Demers.etal-PRD} have the same form as the Noether
charge entropy (\ref{eq:SNC}).

These conclusions raise a related question: {\it Why the entropy of
a black-hole, and not (neutron) star, has the same dimensional form
as the classical gravity action?}  Classically, stars and
black-holes are described by spherically symmetric solutions of
gravity and matter action (\ref{eq:SEH},\ref{eq:SGen}).  However, it
is the existence of the event-horizon which distinguishes
black-holes and stars. Hence, quantum gravity should have a
mechanism to account the existence of the horizon which would imply
that the semi-classical entropy of black-holes, and not stars, has
the same dimensional form as gravity action.

This raises another question: {\it Is there one universal  feature
which is common to the microscopic theory which  distinguishes
black-hole and star?}  Entanglement, the quantum correlation that
exist between subsystems of a quantum system, is a feature of
quantum system. The presence of the event-horizon gives rise to
natural emergence of entanglement entropy
\cite{1986-Bombelli.etal-PRD,1993-Srednicki-PRL} and, hence,
distinguishing the entropy associated to black-holes and stars.

Interestingly, entanglement provides natural explanation for the
area-dependence of black-hole entropy. For a bipartite system in a
pure state, tracing over given subsystem and its complementary
system yield identical entanglement entropies
\cite{2010-Eisert.etal-RMP}. As shown in \cite{2008-Das.etal-PRD},
the interaction terms across the boundary contribute significantly
to the entanglement entropy, hence, entanglement entropy is a
function of the boundary [$\SEN \propto F(A)$] which in the case of
black-holes is the event-horizon.  It is now known that, for fields
in (i) {\sl vacuum}: $F(A) = A$
\cite{1986-Bombelli.etal-PRD,1993-Srednicki-PRL}, (ii) {\sl excited
states}: $F(A) = c_0 A + c_1/A^{\mu}$ ($c_0, c_1, \mu$ are positive
real numbers) \cite{2008-Das.etal-PRD,2008-Das.etal-Arx}.

The central thesis in this article has been to put together some
pieces of the {\it jig-saw puzzle} which one encounters in obtaining
a microscopic description of black-hole entropy. Starting from the
observation that the entropy of black-hole has the same dimensional
dependence as that of classical gravity action, we have shown that
it is plausible to differentiate between different approaches.
Interestingly, this provides a plausible understanding for the
connection between Einstein's equations and thermodynamic equation
of state. We also have provided arguments as to how entanglement
provides a natural framework to understand black-hole
thermodynamics.

There are several conceptual issues which are unresolved: What makes
gravity special that the dimensional analysis of the gravity action
almost directly implies the maximum entropy the gravitational object
can have? How to show from fundamental principle that the above
observation indeed is the key to rewrite Einstein equation as
thermodynamics of space-time? These are currently under investigation.

\section*{Acknowledgements}
The author thanks Saurya Das for many useful debates and comments.
This work is supported by the Department of Science and Technology,
Government of India through Ramanujan fellowship.

\providecommand{\href}[2]{#2}\begingroup\raggedright\endgroup

\end{document}